\documentclass[twocolumn]{aastex6}
\usepackage{booktabs}
\usepackage[]{graphicx} % http://ctan.org/pkg/graphicx
\newcommand{\Spitzer}{{\it Spitzer}}
\newcommand{\HST}{{\it HST}}

\newcommand{\hii}{\mbox{H{\sc ii}}}

\newcommand{\cOri}{{\it c~}Ori}

\begin{document}
%\title{Proplyds around a B1 star - $\it \MakeLowercase{c}$~Orionis in NGC 1977}
\title{Proplyds around a B1 star - 42~Orionis in NGC 1977}
%
%
%% Use \author, \affil, plus the \and command to format author and affiliation 
%% information.  If done correctly the peer review system will be able to
%% automatically put the author and affiliation information from the manuscript
%% and save the corresponding author the trouble of entering it by hand.
%%
%% The \affil should be used to document primary affiliations and the
%% \altaffil should be used for secondary affiliations, titles, or email.
%% Authors with the same affiliation can be grouped in a single
\author{Jinyoung Serena Kim\altaffilmark{1, 2, 3}, Cathie J. Clarke\altaffilmark{4}, 
Min Fang\altaffilmark{1} and Stefano Facchini\altaffilmark{5}}
%
%\author{Jinyoung Serena Kim\altaffilmark{1, 2}}
\altaffiltext{1}{Steward Observatory, University of Arizona, 933 N. Cherry Ave, Tucson, AZ 85721-0065, USA}
\altaffiltext{2}{serena@as.arizona.edu}
\altaffiltext{3}{visitor, Institute of Astronomy, University of Cambridge, Madingley Road, Cambridge CB3 0HA, United Kingdom}
\altaffiltext{4}{Institute of Astronomy, University of Cambridge, Madingley Road, Cambridge CB3 0HA, United Kingdom}
\altaffiltext{5}{Max-Planck-Institut f\"ur Extraterrestrische Physik, Giessenbachstrasse 1, 85748 Garching, Germany}
%\affil{Steward Observatory, University of Arizona, 933 N. Cherry Ave, Tucson, AZ 85721-0065, USA, serena@as.arizona.edu}
%
%\author{Cathie Clarke\altaffilmark{3}}
%\affil{Institute of Astronomy, University of Cambridge, Madingley Road, Cambridge CB3 0HA, United Kingdom}
%
%\and
%\author{Min Fang\altaffilmark{1}}
%\affil{Steward Observatory, University of Arizona, 933 N. Cherry Ave, Tucson, AZ 85721-0065, USA}
%
%\end
%
%% Use the \and command so offset the last author.
%\and
%\author{Min Fang\altaffilmark{1}}
%\affil{Steward Observatory, University of Arizona}
%\altaffiltext{2}{serena@as.arizona.edu}
%\altaffiltext{3}{cclarke@ast.cam.ac.uk}
%\altaffiltext{4}{mfang@email.arizona.edu}
%\altafiiltext{5}{facchini@mpe.mpg.de}
\begin{abstract}
We present the discovery of seven new proplyds (i.e. sources surrounded
by cometary H$\alpha$ emission characteristic of offset ionization
fronts) in NGC 1977, located about 30' north of the
Orion Nebula Cluster at a  distance of $\sim400$~pc.
Each of these proplyds are situated at projected distances  
$0.04-0.27$~pc from the B1V star 42 Orionis (\cOri), which is the main 
source of UV photons in the region. 
In all cases the ionization fronts of the proplyds are clearly 
pointing toward the common ionizing source, 42~Ori,
and 6 of the 7 proplyds clearly show tails pointing away from it. 
These are the first proplyds to be found around a B star, 
with previously known examples instead
being located around O stars, including those in the Orion Nebula Cluster 
around $\theta^1$~Ori~C. The radii of the offset ionization fronts
in our proplyds are between $\sim200$ and $550$~AU; two objects
also contain clearly resolved central sources that we associate
with disks of radii $50-70 $ AU. The estimated strength of the
FUV radiation field impinging on the proplyds is around $10-30$ times
less than that incident on the classic proplyds in the Orion Nebula
Cluster.  We show that the observed proplyd sizes
are however consistent with recent models for FUV photoevaporation in
relatively weak FUV radiation fields.
\end{abstract}
%% automatically renumber them
\section{Introduction} \label{sec:intro}
The star formation environment is likely to affect the evolution of 
protostellar and protoplanetary disks.  Thermally driven winds, 
heated by ultraviolet radiation from massive stars, can shorten
the lifetime of disks around neighboring low mass stars with potentially 
important implications for  giant and icy planet formation.
Observational support for disk destruction in strongly
irradiated environments is provided by the reduction of
disk fraction in the vicinity of O stars in clusters such as
NGC~6611 \citep{guarcello07, guarcello09,guarcello10}, Pismis24 \citep{fang12}
NGC~2244 \citep{balog07}, the Arches Cluster \citep{stolte10}, and
Cygnus~OB2 \citep{guarcello16}, 
although other studies (e.g. \citet{roccatagliata11} in IC~1795 and 
\citet{richert15}  in NGC~6611) have {\it not} found such a decline. 
Likewise \citet{mann14} (see also \citet{mann10}) have shown
that the dust component of disks tends to be less massive in the
immediate vicinity of the dominant O star in the Orion Nebula Cluster (ONC), 
$\theta^1$Ori~C (though \citet{mann15} found no such trend in another 
cluster containing O stars, NGC~2024). 
Finally, it should be noted that although a decline of disk fraction
in crowded areas could in principle be instead attributed to
dynamical interactions (e.g. \citet{pfalzner04,protegies16})
it can be shown that for  a normal IMF, photoevaporation becomes 
an important disk destruction mechanism at substantially lower 
densities than dynamical encounters  \citep{scally01}.

However, the most dramatic evidence of disk destruction is provided by
the large number of {\it proplyds}, cometary objects imaged in H$\alpha$ 
in the vicinity of  $\theta^1$Ori~C in the ONC \citep{odell93,bally00}\footnote{Note 
that the term was initially applied to any disk rendered visible by its proximity 
to an \hii\ region but we here adopt the more restricted definition above which 
has become common usage.}.
The sizes (few hundred AU) and morphologies of proplyds are well 
explained by a model in which FUV radiation from $\theta^1$Ori~C drives a neutral
disk wind; the bright cometary feature then results from the interaction of 
ionizing radiation (also from $\theta^1$Ori~C) with this neutral wind 
\citep{johnstone98}.  Radio free-free measurements of the 
ONC proplyds imply mass loss rates of $\sim 10^{-7} M_\odot$ yr$^{-1}$
\citep{churchwell87}, which would result in extremely
short disk lifetimes. It is therefore unsurprising that objects
experiencing such extreme photoevaporation are observed rather
rarely, with relatively small samples being identified
in Carina \citep{smith03}, Pismis 24 \citep{fang12}, 
NGC3603 \citep{brandner00} and CygOB2 \citep{wright12}; 
in the latter two environments the `giant' proplyds 
(on a scale of $10^4-10^5$ AU) are not necessarily derived from disk
photoevaporation \citep{sahai12a,sahai12b}, though see also \citet{guarcello14} . 

To date proplyds have only been detected around stars of spectral type O, 
where the high mass loss rates both measured and predicted imply 
that this should be a short-lived evolutionary stage.  
Investigating the formation potential of proplyds around O stars, 
 \citet{storzer99} also argued that proplyds should only be 
 detectable in the very close vicinity of such objects, 
 where the FUV field exceeds $5 \times 10^4 G_0$ since their calculations 
 implied that at lower G$_0$, neutral wind driving would be
 too weak to push the ionization front away from the disk, given
 the strong ionizing flux produced by O stars (here G$_0$ is the local
 FUV interstellar field ($1.6\times10^{-3} {\rm erg~cm}^{-2}s^{-1}$). 
 As noted by \citet{storzer99}, this lower limit on FUV
 field strength required for proplyd production is however sensitive to
 stellar spectral type since this controls the relative strength of the 
 FUV and ionizing radiation fields.

More recent studies have emphasized that
significant winds can be driven at considerably lower $G_0$ values
\citep{adams04, facchini16}. The radius of a proplyd produced
by interaction between this neutral wind and the B star's ionizing
luminosity could then be used to {\it measure} the mass loss
rates at lower $G_0$, a quantity that is of great importance in assessing
the significance of photoevaporation in a wide range of star forming
environments. Although the lower ionizing flux of the B star
would produce structures of lower surface brightness than in the
O star case, it would also imply spatially larger proplyds for
a given wind rate, rendering them potentially more detectable than in the
O star case. Moreover, the fact that lower wind mass loss rates are expected
implies that such structures would be longer-lived than their counterparts
around O stars and hence more abundant  on those grounds. 

\begin{figure}[h]
\figurenum{1}
\includegraphics[scale=0.45]{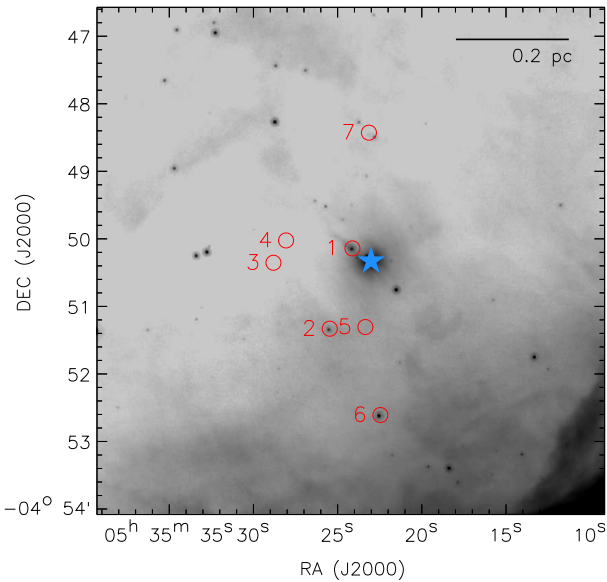} 
\caption{\Spitzer\ 8\micron\ image of NGC 1977 centered at 42~Ori (blue filled star at the center). 
Locations of proplyds are shown as open circles with labels. 
All 7 proplyds (KCFF~1-7) are within 0.3~$pc$ distance from 42~Ori.\label{fig:allproplyds}}
\end{figure}

Here we present seven proplyds discovered around 42~Ori (\cOri,  HD37018, B1V) 
in NGC~1977, an \hii\ region located at $\sim$30\arcmin\ north of 
ONC at $\sim$414~pc distance \citep{menten07}.
There is no O star in the region, but NGC~1977 contains three young B stars 
and at least $\sim$170 young stellar objects \citep{peterson08}.  
42~Ori has the earliest spectral type, and  is the major source for 
ionizing photons in NGC~1977.  
An irradiated disk near 42~Ori  has been detected by \citet{bally12} 
in the \HST\ image using H$\alpha$ filter (F658N). They identified 
a bent protostellar jet HH1064 from Parengo~2042 (the Spindle) 
in NGC~1977 with numerous bow shock features.  
They argue that the arc feature in the H$\alpha$ Spindle 
is centered on the star and its brightened side of the arc is
facing toward 42~Ori, suggesting that it may be a proplyd. 
The seven proplyds that we describe here (Figure~\ref{fig:allproplyds}) were  
discovered in archival \Spitzer\ and \HST\ images; we will discuss the implication of  finding {\it proplyds}
around a B star and the wider implications for disk clearing in UV environments.
%
%%%%%%%%%%%%%%%%%%%%%%%%%%%%%%%%%%%%%
\section{\Spitzer\ \& \HST\ Archival Data} \label{sec:obs}
\begin{figure}[ht]
\figurenum{2}
\includegraphics[scale=0.36]{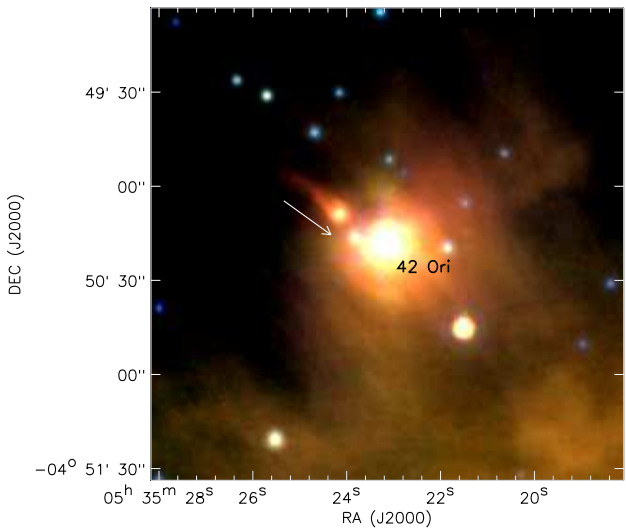}
\caption{Proplyd KCFF 1 identified in the \Spitzer\ images using 3.6\micron (blue), 5.8\micron\ (green), 
and 8.0\micron\ (red). The white arrow shows the direction toward the ionizing source, 42 Ori. \label{fig:KCFF1}}
\end{figure}

%
%\subsection{{\Spitzer} IRAC images}
We used the \Spitzer\ Space Telescope/IRAC (3.6, 4.5, 5.8, 8.0\micron) 
archival data for the detection of a dusty proplyd, KCFF~1.  
All four IRAC bands show clear detection of the central source and 
its tail pointing radially away from the B1 star, 42~Ori (Figure~\ref{fig:KCFF1}).  
The mosaic images were obtained from the \Spitzer\ Heritage Archive
(SHA)\footnote{\url{http://sha.ipac.caltech.edu/applications/Spitzer/SHA/}}.
The images were processed by the pipeline version S18.25.0, and 
the mosaic images were made using MOPEX version 18.5.4 
and super mosaic pipeline version 2.0. The final mosaic image has
resolution of $0\farcs6$.
The median exposure times (seconds per pixel) of the images are   
52 sec for long exposures and 2 sec for short exposures.   

\begin{figure*}[tbp]
\figurenum{3}
\centering \includegraphics[scale=0.9]{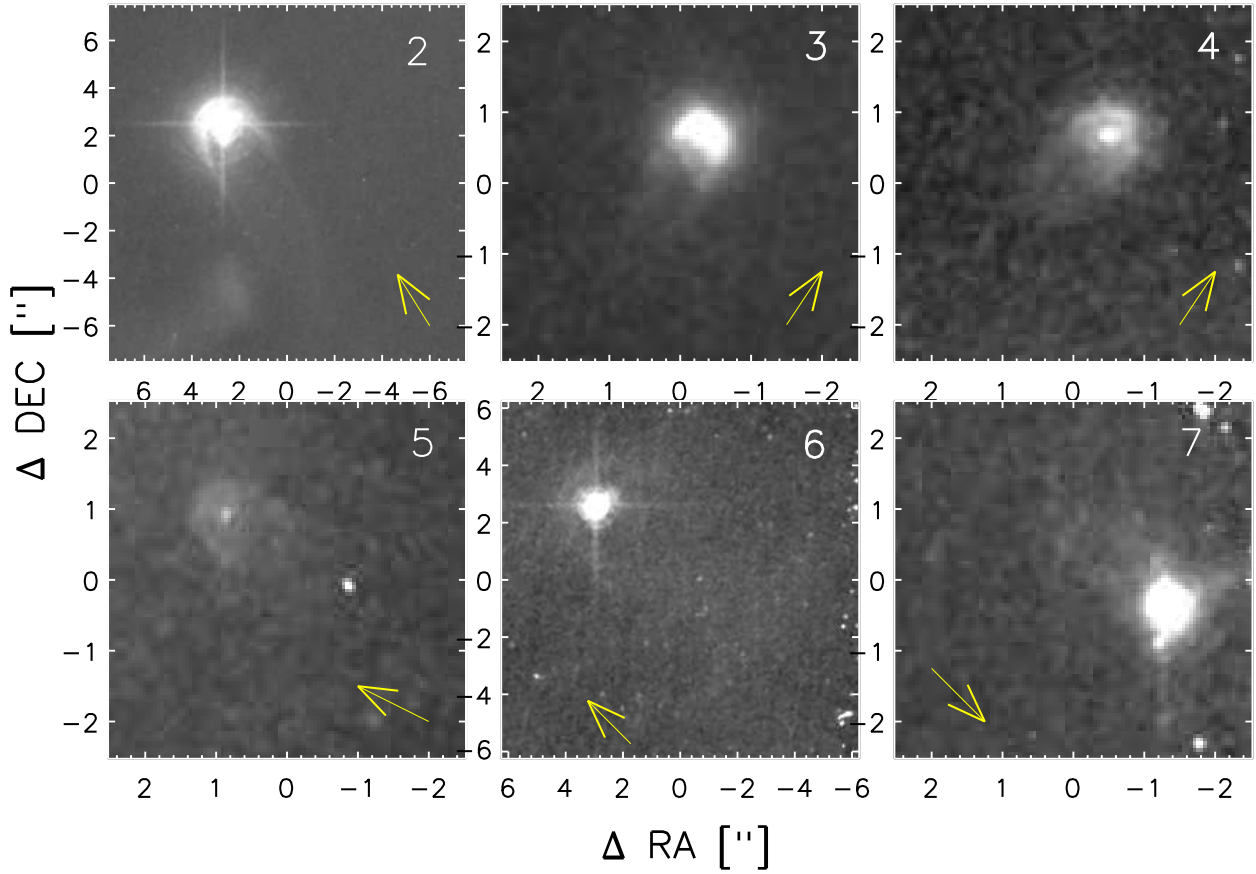}
\caption{Proplyds KCFF 2, 3, 4, 5, 6, \& 7 identified in the \HST/ACS image (F658N). Yellow arrows 
indicate direction toward the B1 star, 42~Ori.\label{fig:6proplyds}}
\end{figure*}

%\subsection{{\it HST} images}
We used a set of archival data of {\it the Hubble Space Telescope} 
(HST)/the Advanced Camera for Surveys (ACS)
to identify proplyds KCFF~2-7 (Figure~\ref{fig:6proplyds}). 
The data were obtained from the MAST archive.   
The \HST\ ACS/WFC images was observed on November 12, 2010 and November 14, 2011 
(PI: Bally, proposal ID 12250, cycle 18) using H$\alpha$ (F658N) filter with
2460 and 2510 second exposure times. The F658N narrow band filter transmits 
both H$\alpha$ and and [N II] lines.   The observational details
were discussed in \citet{bally12}. 
The final mosaic images have pixel sizes of $0\farcs05$\footnote{Based on observations made with the NASA/ESA Hubble Space Telescope, obtained from the data archive at the Space Telescope Science Institute. STScI is operated by the Association of Universities for Research in Astronomy, Inc. under NASA contract NAS 5-26555.}.

\begin{figure*}[tbp]
\figurenum{4}
\centering \includegraphics[scale=0.6]{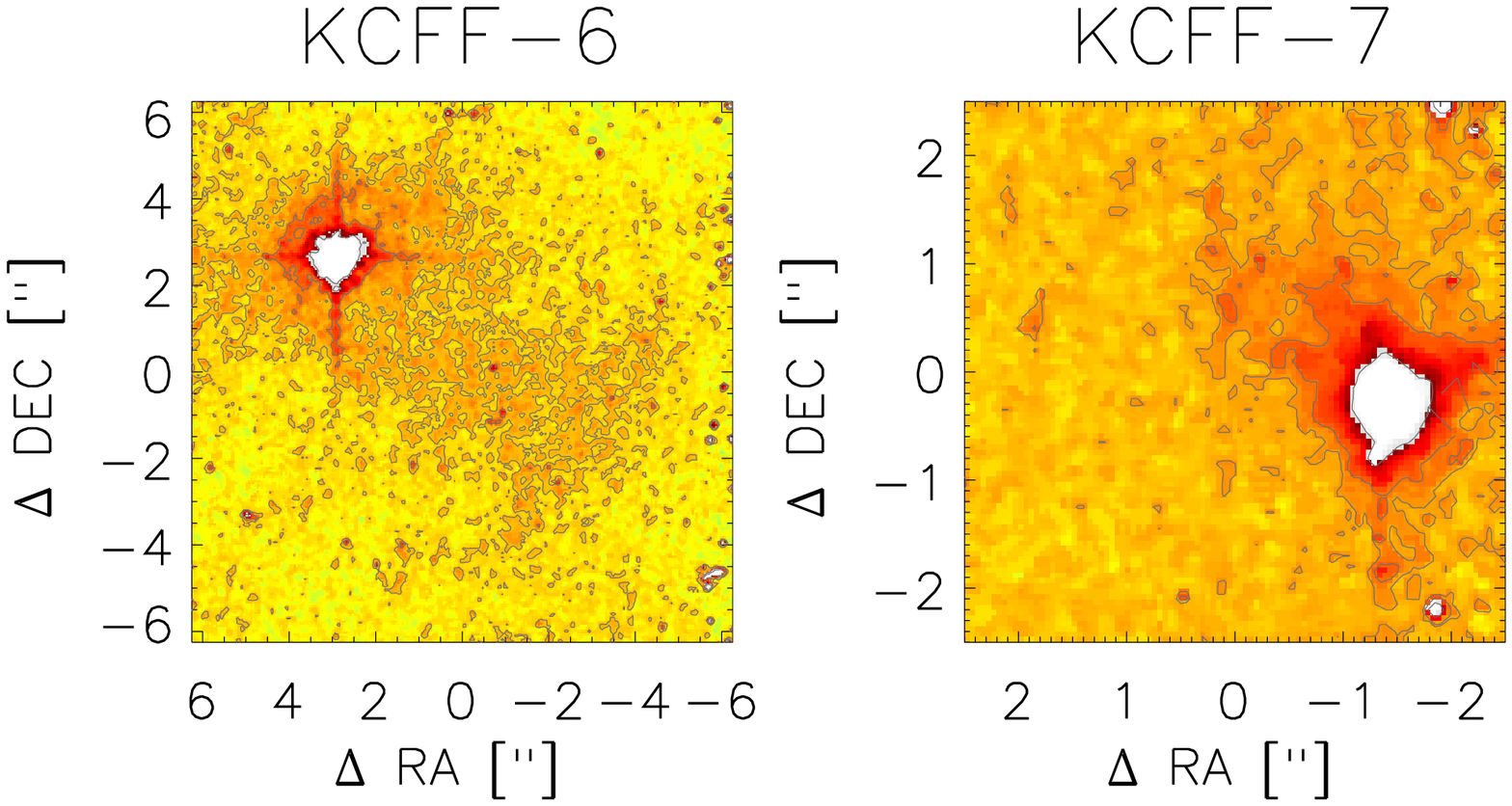}
\caption{The \HST/ACS images (F658N) of the proplyds KCFF  6, \& 7. 
We show the two faint proplyds, KCFF 6 and 7, in color scale with contours in gray 
at 60\%, 67\%, and 90\% of the maximum flux level.\label{fig:KCFF67}}
\end{figure*}

%% "Perhaps a missing \item?"

\section{Proplyds around 42 Ori}
We present a total of 7 new proplyds in NGC 1977 in Table~\ref{tab:coord}
(also see Figure~\ref{fig:allproplyds}).
The KCFF source ID number (1$^{st}$ column) is used when 
we discuss individual sources here, and we also assign names 
for the proplyds based on their coordinates (2$^{nd}$ column of Table ~\ref{tab:coord})
similar to the designation given for proplyds in the ONC \citep{odell98}.  
We use the last three digits in {\it right ascension} (J2000) ($s$\farcs$ss$ after 5$^h$35$^m$20$^s$) 
and the last five digits of {\it declination} (J2000) ($m$:$ss\farcs$$ss$)
to assign names (Table~\ref{tab:coord}, $2^{nd}$ column) for the 
coordinate-based names of the 7 proplyds. 
For example proplyd KCFF~1 with coordinates 5:35:24.142, -4:50:09.21 is named as 
{\it 414-50092}.% in order not to be confused with proplyds in the ONC.

% KCFF-1
The closest proplyd KCFF~1 (414-50092, Figure~\ref{fig:KCFF1}) is discovered in 
the \Spitzer\ images at a distance of $\sim$7000~AU from the B1 star
with its dusty tail evaporating away from 42~Ori.  
Sizes and distances from 42~Ori  are calculated assuming that the distance 
to NGC 1977 is $400~pc$ (throughout the paper). 
This source is detected in all four IRAC bands and in MIPS 24\micron, but not in 
the \HST\ image, because the HST survey area did not cover KCFF-1. 
The central source is an M3.5-M4 star, and its dusty tail is about 
20$\arcsec$ ($\sim$8000~AU) long.

%KCFF2-7
Six other proplyds (KCFF~2$-$7) are identified in the \HST\ ACS/WFC image (Figure~\ref{fig:6proplyds}).  
\citet{bally12} presented Parengo 2042, the {\it Spindle},  residing  in a large proplyd.  
However we do not discuss this source in this paper, since its tail is not clearly 
visible in the \HST\ image alone. 
We measure the radius of the ionization front ($r_{IF}$) and disk radius ($r_d$) 
in a  manner similar to section 3.2 in \citet{vincente05}, where they measure IF chord diameter. 
For  the brightest proplyd, KCFF-2 (551-51201), we fit a circle to $\sim$30\%\ 
higher than the background level, and for other fainter sources we 
fit a circle to $\sim$10\%\ higher than the background using contour, 
radial profile, and cross section analyses. 
The measurement uncertainties in size estimation are 
about half a pixel to a pixel ($\sim0\farcs025 - 0\farcs05$, $\sim$10-20~AU);
this is introduced when we fit a circle to the 10\% or 30\% above the average
background level, and also because of slight departure from a circular 
shape of the proplyd's ionization front (IF).  
The IF radius is estimated as the radius of the circle fitting the IF chord.

All proplyds are found within $0.3$~pc distance from 42~Ori
%(5:35:23.080 -04:50:18.60), 
with their IF pointing toward 42~Ori and their tails pointing radially away from it
(see Fig~\ref{fig:KCFF1} \& \ref{fig:6proplyds}).
KCFF~2 has a bright thick ionization front and a bright central source
with $r_{IF}\sim1\farcs35$~(540~AU) and a central source with radius$\sim0\farcs35$(140~AU).
We do not list this central source as a disk source, since the central source
does not have clear disk morphology.
KCFF~3 (881-50220) has $r_{IF}\sim0\farcs4~$(160AU), but the central source is not detected
while the ionizing front is very bright. 

KCFF~4 \& 5 have resolved central sources. The central source of KCFF~4 (808-50020)
is an illuminated disk {\bf or quasi spherical material} with $r_d\sim0\farcs17$ ($\sim70$AU),   
which is slightly asymmetric with the semi-major axis,
located  inside the proplyd with $r_{IF}\sim0\farcs46$ (184~AU).
The central source of KCFF~5 (338-51180) is smaller than KCFF-4 
with $r_{d}\sim0\farcs12~(48AU)$ and $r_{IF}\sim0\farcs49$~(196~AU).
The size of $r_{IF}$ for these two sources are similar.{\footnote{Note that 
KCFF~4 and KCFF~5 are unique among proplyds
in that their disks are observed in H$\alpha$ {\it emission} as opposed
to the dark (silhouette) disks seen in other regions. While this is
readily explicable in terms of the fainter background in NGC 1977, it still
leaves the open question of how ionizing photons are able to reach these disks.}.

KCFF~3, 4, and 5 are undetected in the {\it Spitzer} photometry catalog 
from \citet{megeath12},  which would suggest that these sources are very 
low mass objects. 
The \Spitzer\ catalog can detect very low mass objects, 
down to brown dwarfs mass objects, because their \Spitzer\ IRAC band~1 (3.6\micron) 
data go as deep as $\sim16~mag$ with 10$\sigma$ detection in Orion.
The non-detection of these sources in the Spitzer IRAC band 1 
would equate to an upper mass limit of around $15$ Jupiter masses according 
to the evolutionary models of \citet{baraffe15}. 
Partial obscuration of the central object by the disk would, however, imply 
that  the mass was considerably greater than this. 
We incline towards this explanation on the grounds that the disk masses and 
consequent disk lifetimes against photoevaporation would otherwise be very short.
Note that by invoking obscuration we are requiring some quasi-spherical 
distribution in the vicinity of the star, or an inclined edge-on disk, which may or 
may not impact our interpretation of the $\sim$50 AU scale structures 
within KCFF~4 and 5 as being disks.

KCFF~6 (252-52365) has a bright central object with low surface brightness 
ionization front with $r_{IF}\sim292$~AU.  
The ionization front size appears to be larger than other proplyds, except KCFF~2.   
The central source  harbors an object with  $T_{eff}$ of 3328~K \citep{dario16}.    
The ionization front is faint, but clearly shows a half-circular morphology 
as a proplyd, but we note that its tail is extremely faint (Figure~\ref{fig:KCFF67})
KCFF~7 (313-48277) has a similar $r_{IF}$ size and a faint tail.  The central
source has the highest $T_{eff}$ (3847~K) among the 7 proplyds in Table~\ref{tab:coord}. 
%The central disk of this proplyd is not detected. 

%\floattable
\begin{deluxetable*}{ccccccccccc}
%\newpage
\tablecaption{Properties of proplyds in vicinity of 42~Ori.\label{tab:coord}}
\tablecolumns{10}
\tablenum{1}
\tablewidth{12pt}
\tablehead{
\colhead{Source ID} & \colhead{Name} & 
\colhead{RA (J2000.0)} & \colhead{DEC (J2000.0)} & \colhead{T$_{eff}$} & 
\multicolumn{2}{c}{distance to 42 Ori} & 
\multicolumn{2}{c}{$r_{IF}$\tablenotemark{d}} & 
\multicolumn{2}{c}{$r_d$\tablenotemark{d}} \\  
\cmidrule{6-7}   \cmidrule{8-9}  \cmidrule{10-11} 
\colhead{KCFF\#} & \colhead{}  &
\colhead{hh mm ss.sss} &\colhead{$\degr~\arcmin~\arcsec$}  & 
\colhead{(K)} & \colhead{(\arcsec)} & \colhead{($pc$)\tablenotemark{a}} &
\colhead{(\arcsec)} & \colhead{($AU$)\tablenotemark{e}}  &  
\colhead{(\arcsec)} & \colhead{($AU$)\tablenotemark{e}} 
}
\startdata
1  &  414-50092  & 5 35 24.142   &    -4 50 09.21  &  3243\tablenotemark{b} &  17.59   &   0.036   & \nodata &\nodata & \nodata  & \nodata\\  
2  &  551-51201  & 5 35 25.505   &    -4 51 20.11  &  3630\tablenotemark{c}  &  70.72   &   0.138   &   1.35  & 540 & \nodata & \nodata \\ 
3  &  881-50220  & 5 35 28.812   &    -4 50 22.04  &  \nodata                          &  84.65   &   0.166   &  0.40  &  160 & \nodata & \nodata \\
4  &  808-50020  & 5 35 28.076   &    -4 50 02.06  &  \nodata                          &  75.45   &   0.148   &  0.46  &  184 & 0.17 & 68   \\
5  &  338-51180  & 5 35 23.381   &    -4 51 18.06  &  \nodata                          &  59.42   &   0.116   &  0.49  &  196  & 0.12 & 48  \\
6  &  252-52365  & 5 35 22.522   &    -4 52 36.56  &  3328\tablenotemark{b}  & 138.14  &   0.268   &  0.73  & 292 & \nodata  & \nodata  \\
7  &  313-48277  & 5 35 23.134   &    -4 48 27.69  & 3847\tablenotemark{c}   & 111.05   &   0.215   &  0.56  & 224 & \nodata  & \nodata \\ 
\enddata
%
%\tablecomments{}
\tablenotetext{a}{Projected distances from 42~Ori to proplyds. We use the distance of NGC 1977 to be 400pc in this work.}
\tablenotetext{b}{\citet{dario16}}
\tablenotetext{c}{Fang et al. 2016, in prep.}
\tablenotetext{d}{Uncertainty of measuring sizes of ionization front and central disk size ranges about $0\farcs025 - 0\farcs05$.}
\tablenotetext{e}{Calculations using $d\sim400~pc$ to NGC 1977.}
%$r_d$ of this source is $\sim0\farcs3-0\farcs4$.}
\end{deluxetable*}

%%%%%%%%%%%%%%%%%%%%%%%%%%%%%%%%%%%%%
%
\section{Modeling} \label{sec:model}
In order to estimate the expected size of proplyds in this environment
(specifically the distance between the center of the proplyd
source and the offset ionization front) we need to
i)  estimate  the ionizing flux from the neighboring B star,  
ii)  estimate the expected mass loss rate in the neutral wind from 
the proplyd and iii) impose a condition of ionization balance in the
ionized flow close to the  ionization front. This approach
can be applied whatever the mechanism driving the
neutral wind \citep{clarke15}. Here we follow
\citet{johnstone98} in assuming that this is driven by the FUV
flux from the same neighboring massive star which also
provides the ionizing photon source. We however differ from
\citet {johnstone98} in that we use new calculations of photoevaporative
mass loss in regions of relatively low FUV fields \citep{facchini16},  
noting that the observed proplyds in NGC 1977 are exposed to 
a FUV field that is less than that irradiating the well studied
proplyds in the ONC.

All proplyds here (KCFF~1$-$7) have  structures with tails pointing
radially away from 42~Ori. It is thus reasonable
to associate the ionization source with this star; indeed the relatively
close proximity of these sources to what is the earliest
type star in the region strengthens this expectation. We estimate
the stellar mass from the B1V spectral type as around $10 M_\odot$
(e.g., \citet{lorenz05,lorenzo16})
and obtain ionizing photon outputs and FUV luminosities of
$10^{45}$ s$^{-1}$ and $2 \times 10^{37}$ erg s$^{-1}$ from
\citet{diazmiller98} and \citet{armitage00} respectively.  We can then
obtain a maximum  FUV flux of $\sim 3000~G_0$ in the
vicinity of the proplyds. This maximum  is obtained by neglecting
dust extinction between the B star and the proplyd and by
setting the distance between star and proplyd to be its separation 
on the sky $\sim0.2-0.3$~pc. 
\citet{storzer99} showed  that 
the creation of proplyds around O stars requires a 
minimum FUV flux of $5 \times 10^4~G_0$.  We here re-examine this 
issue using the lower ionizing fluxes of B stars and more recent models of 
neutral winds from protoplanetary disks in mild FUV environments by 
\citet{facchini16}  (see also the pilot solutions of  \citet{adams04}). 
The mass loss rate in this 
regime is a sensitive function of the disk outer radius and also
depends somewhat on the effect of grain growth in modifying the FUV
opacity in the wind. As an example, we take the cases of moderate
grain growth (maximum grain size of $3.5 \mu$m) and disk radii of
$40$ and $50$ AU, for which the mass loss rates for an FUV
field of $3000~G_0 $ are $10^{-9}$ and $10^{-8} M_\odot$ yr$^{-1}$ respectively 
(see Figure 12 of {\citet{facchini16}). Combining equations (6) and (10) of 
\citet{johnstone98} in order to remove their dependence on the explicit
formula for FUV driven mass loss rate as a function of system
parameters) we obtain $r_{IF}=  1200 AU \Phi_{45}^{-1/3} \dot M_{-8}^{2/3} d_{pc}^{2/3}$
where $\Phi_{45}= \Phi/10^{45} s^{-1}$($\Phi$ being the stellar
ionizing luminosity taking into account possible absorption by dust
and the background nebula), 
$\dot M_{-8}= \dot M/ 10^{-8} M_\odot$yr$^{-1}$ and $d_{pc}$ 
is the distance to the ionizing source in parsecs.
Adopting a distance between the proplyds and the B1 star of $\sim0.2~pc$
and the mass loss rates given above, we obtain $r_{IF}$=$70$ AU  and
$400$ AU for disk radii of $40$ and $50$ AU respectively; $r_{IF}$ values
that are higher by a factor of a few are obtained in the case of
dust growth to mm sizes (see right hand panel of Figure 12, \citet{facchini16}).

We thus see that ionization fronts with offset distances
on  the observed scale ($\sim200$~AU)  
are to be expected given the distances of the proplyds from the
B1 star in NGC~1977.{\footnote{Note that an ionization front with
such an offset is expected from {\it any} mechanism that drives a
neutral wind of this magnitude; such a structure could also thus
be explained by wind produced by internal X-ray photoevaporation
in the case of an X-ray luminosity $\sim 10^{30}$ erg s$^{-1}$
\citep{clarke15}; such an explanation is however unnecessary given that
externally driven FUV photoevaporation can produce such a mass loss rate.}
%
%%%%%%%%%%%%%%%%%%%%%%%%%%%%%%%%%%%%%%%%%%
\section{Conclusion}

We have presented 7 new proplyds around a B1 star, 42~Ori, in NGC~1977, 
about 30\arcmin\ north of the Orion Nebula (M42). This is, to our knowledge, 
the first time that proplyds (i.e. imaged structures showing clear evidence of
external photoevaporation) have been detected in the neighborhood of
stars of later spectral type than O type. This discovery therefore opens up
the possibility of testing theories of FUV photoevaporation in much weaker
FUV background fields than has been possible hitherto (the estimated
FUV background at the location of these new proplyds is $\sim 3000~G_0$, 
more than an order of magnitude less than the estimated fields in the vicinity of the
classical proplyds in the ONC).

Two proplyds (KCFF~3 \& 4) contain bright interior structure on a 
scale of $\sim 50-70$~AU.  We have used the recent models
of \citet{facchini16} to estimate the mass loss rate from such disks 
in radiation fields of $\sim3000~G_0$ and find values in the range
$10^{-9}-10^{-8}~M_\odot$yr$^{-1}$. Such rates are comparable
with typical accretion rates in T Tauri stars, and therefore suggest 
that external photoevaporation will be a major player in the evolution 
of such disks.  We have already noted that the (lack of) \Spitzer\
detection in KCFF~4 and KCFF~5 might imply very low mass central 
objects in these cases; if so, the low expected mass of associated disks
would in turn imply very short disk depletion timescales. Alternatively
these masses may be substantially under-estimated if the source is
partially obscured by an edge-on disk or quasi-spherical material. 

Given estimates for the ionizing flux from 42~Ori that is incident
on the proplyds, we use these mass loss estimates to predict
the expected radii for offset ionization fronts in these objects
and obtain values of order $100$~AU (the predicted mass loss 
rates and the resulting proplyd radii are a sensitive function of disk
radii, which can be estimated only in a few cases). These predicted
proplyd sizes are in excellent agreement with the scales
of structures seen in the proplyds of 42~Ori.

\acknowledgments
We thank the anonymous referee for the helpful comments.
This work has been partially supported  by  the DISCSIM project, grant agreement 
341137 funded by the European Research Council under ERC-2013-ADG.
This work benefits from the EOS NExSS collaboration.
The HST data presented in this paper were obtained from the Mikulski Archive for Space Telescopes (MAST)
at {\url{https://archive.stsci.edu/hst/}.
STScI is operated by the Association of Universities for Research in Astronomy, Inc., under NASA contract NAS5-26555. Support for MAST for non-HST data is provided by the NASA Office of Space Science via grant NNX09AF08G and by other grants and contracts. 

% 
%%%%%%%%%%%%%%%%%%%%%%%%%%%%%%%%%%%%%%%%%%%


\listofchanges

\end{document}